\documentclass[12pt,aps,prb,showpacs,preprint]{revtex4}   

\usepackage{amsmath}    
\usepackage{graphicx}   
\usepackage{dcolumn}    
\usepackage{bm}

\begin{document}

\title{Spurious causality hiding an action at a distance }
\author{Chun Wa Wong}
 \affiliation{Department of Physics and Astronomy, 
University of California, Los Angeles, CA 90095-1547, USA}   
 \email{cwong@physics.ucla.edu}   

\date{\today}

\begin{abstract}
The Coulomb/Gauss law given in the Maxwell equations describes a spatial relation
between the electric field component $E_\parallel({\bf r}, t)$ and its source 
$\rho({\bf r}, t)$ that is instantaneous, occurring at the same time $t$. 
This instantaneous action at a distance can be hidden by writing it formally 
as the sum of two causal terms. I show here that the causal expression of the 
total electric field ${\bf E}$ suffers from such 
a spurious causality hiding the action at a distance that is $E_\parallel$. 
In fact, $E_\parallel$ remains instantaneous in time even when it is part of
a wave equation. Only the ${\bf E}_\perp$ part is really causal. 
\end{abstract}


\maketitle

\section{The puzzle}

It is well known that far from charges and currents, the electromagnetic (EM) fields 
{\bf E} and {\bf B} satisfy homogeneous vacuum wave equations where the speed of 
light $c$ appears. These wave equations involve only the transverse parts of the EM 
fields. 

The situation is more complicated when the system is close to charges and currents. 
The vacuum wave equations are then inhomogeneous. Most textbooks derive directly 
from the Maxwell equations the inhomogeneous wave equations for these EM fields,
using the Lorenz gauge first introduced by L. V. Lorenz in 1867.\cite{Lorenz67} 
Many textbooks also express the scalar and vector potentials separately as retarded 
solutions relative to their sources, namely the charge and current densities, 
respectively. These expressions were also first used by Lorenz in 
1867.\cite{Lorenz67, McDonald97, Jackson01} 

More recently, Panofsky and Phillips\cite{Panofsky62} and 
Jefimenko\cite{Jefimenko66,Jefimenko89} have given expressions for 
the entire {\bf E} and {\bf B} fields where every term is retarded with respect 
to the charge-current sources. The results are considered ``time-dependent 
generalizations of the Coulomb and Biot-Savart laws.''\cite{Griffiths91, McDonald97, 
Jackson99}

Yet the electric field {\bf E} has an identifiable part that satisfies the original
vacuum Coulomb/Gauss law in the presence of a time-dependent source
\begin{eqnarray}
\bm {\nabla \cdot} {\bf E}({\bf r}, t) = \rho({\bf r}, t)/\epsilon_0
\label{Gauss}
\end{eqnarray}
in SI units. This Maxwell equation states unambiguously that the part of the field 
{\bf E} involved in the equation is nonlocally related to its source $\rho$ in 
space, but it occurs at the same time or instantaneously as the source 
$\rho$.\cite{Wong09, Wong10}

That the result involves only a part of {\bf E} can be made manifest by working
in the Fourier space {\bf k} instead of the real space {\bf r}. The notation and
results of Wong\cite{Wong09} will be used here to avoid duplication. In the Fourier 
space, the Coulomb/Gauss law reads 
\begin{eqnarray}
\tilde{E}_\parallel({\bf k}, t) =  -\frac{i}{\epsilon_0 k}\tilde{\rho}({\bf k}, t).
\label{Epara}
\end{eqnarray}
Thus only $\tilde{E}_\parallel$ is instantaneous in time. 

As shown in Ref. \onlinecite{Wong09}, Eq. (\ref{Epara}) appears twice in the 
Maxwell equations. It can also be obtained a second time as the longitudinal part 
of the Maxwell $\bm{\nabla \times} {\bf B}$ equation 
\begin{eqnarray}
\tilde{\bf J}_\parallel({\bf k}, t)
 + \epsilon_0 \partial_t \tilde{\bf E}_\parallel({\bf k}, t) = 0,
\label{Jpara}
\end{eqnarray}
when use is made of the continuity equation. Here $\partial_t = \partial/\partial t$. 
Equation (\ref{Jpara}) thus provides a counterexample to the claim of 
Ref. \onlinecite{Griffiths91} that the two sources {\bf J} and $\partial_t{\bf E}$ 
of the $\bm{\nabla \times} {\bf B}$ equation are completely independent of each other.

Can the electric field {\bf E} be causal as a whole, and yet instantaneous in its
longitudinal part? It is puzzling how an identifiable part 
${\bf E}_\parallel$ of the electric field that is an action at a distance can 
disappear from plain sight, like the tiger in a magician show. The purpose of 
this paper is to show that the trick is done by writing it as a sum of two causal 
terms, and that ${\bf E}_\parallel$ itself, though cut up and disguised, remains 
instantaneous in time. In Sect. \ref{sect:spurious}, I shall do this by working
directly with electric fields without using scalar and vector potentials in order 
to preserve manifest gauge independence.

The same puzzle does not appear in {\bf B}. With no monopole present,
${\bf B}_\parallel = 0$. Therefore ${\bf B} = {\bf B}_\perp$ is entirely 
transverse and causal. It is of course well defined both mathematically and   
physically. ${\bf E}_\perp$ is intimately related to ${\bf B}_\perp$. They
drive each other forward as an electromagnetic wave through Faraday induction 
and Maxwell displacement. So ${\bf E}_\perp$ too is well defined both 
mathematically and physically. Finally ${\bf E}_\parallel = {\bf E} - {\bf E}_\perp$ 
is well defined mathematically and physically.

There have been at least two previous attempts to explain why the entire {\bf E} 
might be causal.\cite{CTDRG89,Jackson10} Both attempts make use of the causal form 
of the scalar and vector potentials in the Lorenz gauge. We shall show in Sect. 
\ref{sect:potentials} that both $\rho$ and $A_\parallel$ suffer from spurious
causality, the same disease that infects the causal form of $E_\parallel$.

Finally, the Maxwell equations and their Fourier representation are given in 
appendix \ref{app:MaxwellEqs} for the reader's convenience.

\section{Spurious causality in the electric field}
\label{sect:spurious}

Spurious causality can be introduced into the instantaneous action at a distance
that is ${\bf E}_\parallel({\bf r},t)$ by working in the full Fourier space 
$({\bf k}, \omega)$. Equation (\ref{Epara}) can then be written as
\begin{eqnarray}
\tilde{E}_\parallel({\bf k}, \omega) 
=  -\frac{i}{\epsilon_0} \frac{\tilde{\rho}}{k} 
\left( \frac{k^2 - \omega^2/c^2}{k^2 - \omega^2/c^2} \right).
\label{Epara-kw}
\end{eqnarray}
Note that the two introduced factors depend on $c$ and are therefore causal. 
They obviously cancel each other so that the entire expression remains instantaneous 
in time on going back to the time description. 

It is possible to hide this cancellation by using the continuity equation for a 
conserved charge density
\begin{eqnarray} 
0 &=& \frac{d\rho}{dt}  = \partial_t \rho  + {\bf v} \cdot \bm{\nabla}\rho 
\nonumber \\
&=& \partial_t \rho + \bm{\nabla \cdot}{\bf J},
\label{ContinuityEq}
\end{eqnarray}
where ${\bf v} = d{\bf r}/dt$ and ${\bf J} \equiv {\bf v}\rho$. In the Fourier 
space $({\bf k}, \omega)$, the continuity equation takes the more transparent form
\begin{eqnarray} 
\omega \tilde{\rho} = k\tilde{J}_\parallel.
\label{ContinuityEq-kw}
\end{eqnarray}
It shows that only the longitudinal component $J_\parallel$ in space-time is involved 
in charge conservation. The transverse component remains irreducible in its original 
${\bf v}\rho$ form. 

Using the continuity equation, one finds in Eq. (\ref{Epara-kw}) a factor
\begin{eqnarray}
\frac{\tilde{\rho}}{k} (k^2 - \omega^2/c^2 ) 
= \tilde{\rho}k - \tilde{J}_\parallel \omega\epsilon_0\mu_0 .
\label{numerator}
\end{eqnarray}
Hence
\begin{eqnarray}
\tilde{E}_\parallel({\bf k}, \omega) 
&=&  -\frac{i}{\epsilon_0(k^2 - \omega^2/c^2)}
( \tilde{\rho}k - \tilde{J}_\parallel \omega\epsilon_0\mu_0) \nonumber \\
&=& \tilde{E}_{J\parallel{\rm causal}} + \tilde{E}_{J\parallel}.
\label{EparaCausal}
\end{eqnarray}
The final two terms are given the notation used in Ref. \onlinecite{Wong10}. 
Each term is causal, propagating with light speed $c$. 

The causality so introduced is totally spurious, however, because their sum is 
already known to be instantaneous. That is, the introduced causality effects must 
cancel out {\it completely} between the two terms. In fact, we could have used a 
nonphysical and arbitrary gauge velocity $v_g$ here, with exactly the same 
cancellation, as will be done in section \ref{sect:potentials}. This is just 
another way of showing that the causality effects involved are really spurious. 
Doing it with the physical light speed $c$, however, seems to have endowed these 
spurious terms with real physical properties.

Spurious causality also infects the vacuum wave equation for the electric 
field {\bf E} in the presence of sources
\begin{eqnarray}
\left( \nabla^2 - \frac{1}{c^2}\partial_t^2 \right) {\bf E}
= \frac{1}{\epsilon_0 } \bm{\nabla} \rho  + \mu_0 \partial_t {\bf J}.
\label{waveEqE}
\end{eqnarray}
Decomposed into longitudinal and transverse parts, conveniently done in the 
Fourier space $({\bf k}, \omega)$, it gives
\begin{eqnarray}
\left (k^2 - \frac{\omega^2}{c^2} \right)\tilde{\bf E}_\perp 
&=& i\mu_0\omega\tilde{J}_\perp, 
\label{waveEqET} \\
\left (k^2 - \frac{\omega^2}{c^2} \right)\tilde{\bf E}_\parallel 
&=& -i\frac{k\tilde{\rho}}{\epsilon_0} + i\mu_0\omega \tilde{J}_\parallel 
\nonumber \\
&=& -\frac{i}{\epsilon_0} \frac{\tilde{\rho}}{k} 
\left (k^2 - \frac{\omega^2}{c^2} \right).
\label{waveEqEL}
\end{eqnarray}
Use has been made of the identity (\ref{numerator}) to reach the last expression. 
The two factors of $(k^2 - \omega^2/c^2)$ cancel in the wave equation for 
$\tilde{\bf E}_\parallel$, leaving the instantaneous result of Eq. (\ref{Epara}). 
The myth that the entire ${\bf E}$ is causal, a myth passed down by generations 
of textbooks since 1867, can finally be laid to rest.

No rearrangements can change these properties. For example, in the wave equation
(\ref{waveEqET}), one can separate the source current into parts 
$\tilde{J}_\perp = \tilde{J} - \tilde{J}_\parallel$ to give the separation
$\tilde{\bf E}_\perp = \tilde{\bf E}_J - \tilde{\bf E}_{J\parallel}$, as done in 
Ref. \onlinecite{Wong10}. Then
\begin{eqnarray}
{\bf E} = {\bf E}_\perp + {\bf E}_\parallel 
 = ({\bf E}_J - {\bf E}_{J\parallel}) + 
({\bf E}_{J\parallel} + {\bf E}_{J\parallel{\rm causal}}).
\label{Edecompositions}
\end{eqnarray}
The final sum ${\bf E} = {\bf E}_J + {\bf E}_{J\parallel{\rm causal}}$ of two 
nominally causal terms actually contains the instantaneous ${\bf E}_\parallel$
term, but this instantaneous component has been cut up and disguised as
causal. Since the longitudinal/transverse separation is unique, the resulting 
causal/instantaneous separation is also unique and beyond doubt.

To summarize, I have verified the result of Ref. \onlinecite{Wong09} that the 
total electric field is made up of a unique causal transverse part ${\bf E}_\perp$, 
containing physical signals traveling at the physical light speed, and a unique 
longitudinal part ${\bf E}_\parallel$ that is an instantaneous action at a 
distance. 

The method of Ref. \onlinecite{Wong09} is cleaner than the demonstration given 
here because the longitudinal part of Maxwell's displacement current is separated 
from the Maxwell $\bm{\nabla \times} {\bf B}$ equation before the rest of the
equation is combined with the $\bm{\nabla \times} {\bf E}$ equation to give the 
transverse wave equations. In this way, the spurious causality seen in Eq. 
(\ref{waveEqEL}) is entirely avoided. The separated longitudinal displacment 
current can then be combined with the continuity equation to give the instantaneous 
Coulomb/Gauss law a second time, as already discussed in connection with 
Eq. (\ref{Jpara}).

\section{Spurious causality in scalar and vector potentials}
\label{sect:potentials}

In the Fourier space $({\bf k}, \omega)$, the transverse fields can be described
uniquely by a transverse vector potential:
\begin{eqnarray}
\tilde{\bf B}_\perp &=& i {\bf k} \bm{\times} \tilde{\bf A}_\perp, 
\quad {\rm where} \quad \tilde{\bf A}_\perp  
= \frac{i}{k^2} {\bf k} \bm{\times} \tilde{\bf B}_\perp
\label{Aperp} \\
\tilde{\bf E}_\perp &=& i\omega\tilde{\bf A}_\perp. 
\label{Eperp} 
\end{eqnarray}
Like $\tilde{\bf B}_\perp$ and $\tilde{\bf E}_\perp$, $\tilde{\bf A}_\perp$ is
causal, propagating with light speed $c$. In classical electrodynamics, one can 
stop at this point and never run into any redundant gauge degree of freedom. 

However, the idea of a potential, so successful in classical mechanics, was too
good to pass up. So it developed that the instantaneous longitudinal electric field 
was given an alternative gauge-dependent form
\begin{eqnarray}
\tilde{E}_\parallel &=& -\frac{i}{\epsilon_0} \frac{\tilde{\rho}}{k} 
\nonumber \\
&=& -ik\tilde{\Phi} + i\omega\tilde{A}_\parallel,
\label{EparallelPhiA}
\end{eqnarray}
where a single field component is described redundantly by two scalar fields
$\tilde{\Phi}$ and $\tilde{A}_\parallel$. The redundancy has to be removed by a
choice of gauge. For example, velocity gauges are defined by the gauge 
condition\cite{Yang05, Jackson02, Wong09}
\begin{eqnarray}
\tilde{A}^{(vg)}_\parallel = \alpha\frac{\omega}{k} \tilde{\Phi}^{(vg)},
\label{gaugeCond}
\end{eqnarray}
where the gauge parameter $\alpha = 1/v_g^2$ can be written in term of a 
gauge velocity $v_g$. With two algebraic equations (\ref{EparallelPhiA}) and 
(\ref{gaugeCond}) in two unknowns $\tilde{\Phi}, \tilde{A}_\parallel$, one can 
solve for them to get the causal expressions
\begin{eqnarray}
\tilde{\Phi}^{(vg)} 
&=& \frac{\tilde{\rho}}{\tilde{\epsilon}_\parallel (k^2 - \omega^2 /v_g^2)},
\nonumber \\
\tilde{A}^{(vg)}_\parallel 
&=& \frac{\omega}{kv_g^2}
\frac{\tilde{\rho}}{\tilde{\epsilon}_\parallel (k^2 - \omega^2 /v_g^2)}
\nonumber \\
&=& \frac{1}{v_g^2}
\frac{\tilde{J}_\parallel}{\tilde{\epsilon}_\parallel (k^2 - \omega^2 /v_g^2)}.
\label{PhiAparaRho}
\end{eqnarray}
In this way, an action at a distance $\tilde{E}_\parallel$ can be cut up and 
disguised as two causal terms. 

The disguise is most effective when $v_g = c$ is used. Combined with the really 
causal transverse vector potential $\tilde{\bf A}_\perp$, one finds Lorenz's 
causal expressions for $\tilde{\rho}$ and $\tilde{\bf A}$ given in many 
textbooks. These potentials are usually obtained by solving the inhomogeneous
wave equations in the Lorenz gauge for these potentials.

The causality they display is totally spurious, however. When substituted into 
Eq. (\ref{EparallelPhiA}), these causal potential components simply give
\begin{eqnarray}
\tilde{E}_\parallel 
= \frac{k^2}{k^2 - \omega^2/v_g^2}\tilde{E}_\parallel 
- \frac{\omega^2/v_g^2}{k^2 - \omega^2/v_g^2}\tilde{E}_\parallel,
\label{EparallelPhi+A}
\end{eqnarray}
The result shows clearly that the introduced spurious causality is canceled out 
completely by the numerator terms. It matter not a whit if $v_g = c$, as used in 
the Lorenz gauge, or if $v_g \neq c$ and therefore obviously unphysical. 
Only when $v_g \rightarrow \infty$, do we recover the original instantaneous, 
non-causal result, free of any masquerading causal terms. 

The original result is just the result obtained in the Coulomb gauge. It is also
the gauge-independent result that comes directly from the Coulomb/Gauss part of 
the Maxwell equations. So this instantaneous result cannot be avoided. In fact, 
Eq. (\ref{EparallelPhi+A}) has been used in Ref. \onlinecite{Wong09} to verify 
the gauge invariance of $\tilde{E}_\parallel$ when constructed from the 
$\tilde{\rho}$ and $\tilde{A}_\parallel$ of any velocity gauge, including the 
Lorenz gauge. 

This is not to say that Lorenz-gauge expressions for the potentials and {\bf E}
are not practically useful and convenient. By using the same causality everywhere,
they achieve great compactness and ease of handling. Well-known examples include 
the Lienard-Wiechert potential of 1898-1900\cite{Whittaker51} and Jefimenko's 
expression for {\bf E}. The only caveat is that when it comes to interpretation, 
the instantaneous nature of ${\bf E}_\parallel$ should be acknowledged if not made 
explicit.

In conclusion, we have verified the result of Ref. \onlinecite{Wong09} that the 
so-called time-dependent generalizations of the Coulomb and Biot-Savart laws and 
the causal interpretation of $\rho$ and ${\bf A}_\parallel$ are mathematically 
misleading and physically meaningless. 

\appendix
\section{Maxwell equations and its Fourier representation}
\label{app:MaxwellEqs}

For the reader's convenience, we reproduce here the vacuum Maxwell equations 
for electromagnetic fields in space-time $({\bf r}, t)$ in SI units in the 
notation of Jackson (p. 248):\cite{Jackson99}
\begin{eqnarray}
\bm {\nabla \cdot} {\bf E} &=& \rho/\epsilon_0, 
\qquad \, \bm {\nabla \cdot} {\bf B} = 0, 
\label{divEqs} \\
\bm {\nabla \times} {\bf E} &=&  - \partial_t {\bf B}, \quad
\bm {\nabla} \bm{\times} {\bf B} = \mu_0{\bf J} + \partial_t {\bf E}/c^2.
\label{curlEqs}
\end{eqnarray}
Here $\partial_t \equiv \partial/\partial t$, while $\rho = \rho({\bf r},t)$ 
and {\bf J} are charge and current densities, respectively. The vector fields
${\bf E} = {\bf E}({\bf r},t)$ and {\bf B} are assumed to have well-defined 
second space-time derivatives so that wave equations can be constructed.

Suppose these fields and their first and second derivatives in space-time  
have the Fourier representations
\begin{eqnarray}
{\bf E}({\bf r}, t) &=& \int_{-\infty}^\infty \frac{d\omega}{2\pi} 
\int \frac{d^3k}{(2\pi )^3} e^{i{\bf k}\bm{\cdot}{\bf r} - i\omega t} 
\tilde{\bf E}({\bf k}, \omega), 
\label{FR-E} \\
\bm {\nabla \cdot}{\bf E}({\bf r}, t) 
&=& \int_{-\infty}^\infty \frac{d\omega}{2\pi} 
\int \frac{d^3k}{(2\pi )^3} e^{i{\bf k}\bm{\cdot}{\bf r} - i\omega t} 
i{\bf k}\bm{\cdot}\tilde{\bf E}({\bf k}, \omega), 
\label{FR-divE} 
\end{eqnarray}
etc. The Maxwell equations in the Fourier space $({\bf k}, \omega)$ then 
simplify to the algebraic equations
\begin{eqnarray}
i{\bf k}\bm {\cdot} \tilde{\bf E} &=& \tilde{\rho}/\epsilon_0, 
\qquad \; \, i{\bf k}\bm{\cdot} \tilde{\bf B} = 0, 
\label{divEqs-kw} \\
i{\bf k}\bm{\times} \tilde{\bf E} &=&  i\omega \tilde{\bf B}, 
\qquad   i{\bf k}\bm{\times} \tilde{\bf B} 
= \mu_0\tilde{\bf J} - i\omega\tilde{\bf E}/c^2.
\label{curlEqs-kw}
\end{eqnarray}
The Helmholtz theorem that decomposes any vector field $\tilde{\bf E}$ into 
longitudinal and transverse parts is just the BAC identity
\begin{eqnarray}
\tilde{\bf E} = \tilde{\bf E}_\parallel + \tilde{\bf E}_\perp
= {\bf e}_{\bf k} ({\bf e}_{\bf k} \bm{\cdot} \tilde{\bf E}) 
- {\bf e}_{\bf k} \bm{\times} ({\bf e}_{\bf k} \bm{\times} \tilde{\bf E}), 
\label{HelmholtzThm}
\end{eqnarray}
where ${\bf e}_{\bf k} = {\bf k}/k$.

\end{document}